\documentclass[aps,floatfix,showpacs,nofootinbib,showkeys,twocolumn]{revtex4}
\usepackage{amsmath,amsfonts,euscript,amssymb}
\usepackage{colordvi}
\usepackage[T1]{fontenc}
\usepackage[english]{babel}
\usepackage{epsfig}
\usepackage{graphicx}
\newcommand{\be}{\begin{equation}}
\newcommand{\ee}{\end{equation}}
\newcommand{\bea}{\begin{eqnarray}}
\newcommand{\eea}{\end{eqnarray}}
\newcommand{\dd}{\mbox{d}} % \dd instead \d for differentials
\newcommand\nn{\nonumber}

\begin{document}

\title{Production of $\omega\pi^0$ pair in electron-positron annihilation}

%\title{Associated production of $\pi^0$ and $\omega$ mesons
%in electron-positron annihilation}

\author{A.\ B.\ Arbuzov}
\affiliation{Bogoliubov Laboratory of Theoretical Physics,
JINR, Dubna, 141980  Russia}
\affiliation{Department of Higher Mathematics, University Dubna,
Dubna, 141980   Russia}

\author{E.\ A.\ Kuraev, M.\ K.\ Volkov}
\affiliation{Bogoliubov Laboratory of Theoretical Physics,
JINR, Dubna, 141980  Russia}

%\begin{history}
%\received{Day Month Year}
%\revised{Day Month Year}
%\end{history}

\begin{abstract}
The process of electron-positron annihilation into a pair of
$\pi^0$ and $\omega$ mesons is considered in the framework of
the SU(2)$\times$SU(2) Nambu--Jona-Lasinio model. Contributions
of intermediate photons, $\rho(770)$ and $\rho'(1450)$ vector mesons are
taken into account. It is shown that the bulk
of the cross section at energies below 2 GeV is provided by
the process with intermediate  $\rho'(1450)$ state.
The contribution due to single photon and $\rho(770)$ exchange
is in agreement with the vector meson dominance model. 
Numerical results are compared with experimental data.
\end{abstract}

\date{\today}



\maketitle

\section{Introduction}

Studies of the process of associated production of $\pi^0$ and $\omega$ mesons
at colliding electron-positron beams provide interesting information about
meson interactions at low energies. Moreover this channel is one of the contributions
to the total cross section of $e^+e^-$ annihilation into hadrons, which is
required for a precise determination of the hadronic vacuum polarization.

The annihilation into the $\omega\pi^0$ pair
at energies below 2~GeV was studied experimentally at DM2~\cite{Bisello:1990du},
ND~\cite{Dolinsky:1991vq}, SND~\cite{Achasov:2000wy}, and CMD-2~\cite{Akhmetshin:2003ag}.
The same interactions can be also found in the tau lepton decay $\tau\to \pi\omega\nu_{\tau}$
studied at CLEO~II~\cite{Edwards:1999fj}.

For theoretical description of the process under consideration the vector-dominance-like
models were used, see {\it e.g.} Ref.~\cite{Akhmetshin:2003ag}.
To fit the experimental data a set of additional parameters describing contributions of amplitudes
with virtual $\rho(770)$, $\rho'(1450)$ and $\rho''(1700)$ mesons was introduced.
The energy dependence of these parameters was neglected.
Earlier the process of $\rho'\to\omega\pi$ decay was considered within
a relativistically generalized quark model in Ref.~\cite{Gerasimov:1981gj}
and in a non-relativistic quark model~\cite{Close:2002ky}.
In Ref.~\cite{Li:2008xm} the reaction $e^+ e^- \to \omega \pi^0$ was considered
in the vicinity of $\phi$ meson mass region, where the KLOE experimental
data is available~\cite{Ambrosino:2008gb}.
In this paper we will not work specially at this resonance,
so that the region from the threshold up to about 2~GeV c.m.s. energy will be considered
without taking into account the $\phi$ meson contribution.
Recently in Ref.~\cite{Kittimanapun:2008wg} the process was considered
in frames of a non-relativistic quark model.
It is argued there that the process at energies below 2~GeV is dominated by
the two-step process in which the primary quark-antiquark pair forms a $\rho$ meson
in the ground or excited state and then the vector meson decays into $\omega$ and $\pi$.
It is important to note that the studies in
papers~\cite{Akhmetshin:2003ag,Edwards:1999fj,Kittimanapun:2008wg}
concluded that the contribution of the $\rho''(1700)$ to the process is small.
Following the results these works we will neglect the contribution of the amplitude
with intermediate $\rho''(1700)$ meson.
Meanwhile in Ref.~\cite{Achasov:1996gw} it is claimed that for a simultaneous description of
a series of different annihilation and decay processes all three rho meson states should be
taken into account.

In the present paper for the description of the process $e^+e^-\to\omega\pi^0$ we will use the version of the
Nambu-Jona-Lasinio (NJL) model, which allows us to describe both the ground and the first radial-excited
meson states~\cite{Volkov:1996br,Volkov:1996fk,Volkov:1997dd,Volkov:1999yi,Volkov:2001ct}.
Note that for the description of the amplitudes with  virtual photon
and the ground $\rho(770)$ state one can use the standard NJL model~\cite{Volkov:1986zb,Ebert:1985kz,Vogl:1991qt,Klevansky:1992qe,Volkov:1993jw,Ebert:1994mf,Hatsuda:1994pi,Volkov:2006vq}. It worth to note that
for the case of the ground meson states both versions of the NJL model lead to the same results,
see {\it e.g.} Refs.~\cite{Volkov:1997dd,Arbuzov:2010vq}.
In our model it is possible to describe as the transition amplitudes $\gamma^*\to\rho,\rho'$ as well
as the vertexes $\gamma^*,\rho,\rho'\to\pi^0\omega$ without introduction of any additional arbitrary
parameters. Moreover, the description of the vertexes using quark triangle diagram of the anomaly
type allows us to
get the energy dependence of them.

\section{Process Amplitudes}

\begin{figure}
\begin{center}
\special{em:linewidth 0.6pt}
\linethickness{0.6pt}
\begin{picture}(150.00,80.00)
\put(40.00,40.00){\vector(-1,-1){15.0}}
\put(25.00,25.00){\line(-1,-1){15.0}}
\put(28.00,20.00){\makebox(0,0)[cc]{$e^+$}}
\put(10.00,70.00){\vector(1,-1){17.0}}
\put(25.00,55.00){\line(1,-1){15.0}}
\put(28.00,65.00){\makebox(0,0)[cc]{$e^-$}}
\put(42.00,40.00){\oval(4.00,4.00)[t]}
\put(46.00,40.00){\oval(4.00,4.00)[b]}
\put(50.00,40.00){\oval(4.00,4.00)[t]}
\put(54.00,40.00){\oval(4.00,4.00)[b]}
\put(58.00,40.00){\oval(4.00,4.00)[t]}
\put(62.00,40.00){\oval(4.00,4.00)[b]}
\put(66.00,40.00){\oval(4.00,4.00)[t]}
\put(70.00,40.00){\oval(4.00,4.00)[b]}
\put(58.00,47.00){\makebox(0,0)[cc]{$\gamma^*$}}
\put(72.00,40.00){\vector(+1,+1){15.0}}
\put(87.00,55.00){\line(+1,+1){15.0}}
\put(102.00,70.00){\vector(0,-1){33.0}}
\put(113.00,40.00){\makebox(0,0)[cc]{$u,d$}}
\put(102.00,40.00){\line(0,-1){30.0}}
\put(102.00,10.00){\vector(-1,+1){17.0}}
\put(87.00,25.00){\line(-1,+1){15.0}}
\put(102.00,11.00){\line(1,0){30.0}}
\put(102.00,09.00){\line(1,0){30.0}}
\put(125.00,65.00){\makebox(0,0)[cc]{$\omega$}}
\put(102.00,71.00){\line(1,0){30.0}}
\put(102.00,69.00){\line(1,0){30.0}}
\put(125.00,17.00){\makebox(0,0)[cc]{$\pi^0$}}
\put(102.00,70.00){\circle*{3.00}}
\put(102.00,10.00){\circle*{3.00}}
\end{picture}
\end{center}
\caption{The Feynman diagram with photon exchange.
\label{Fig:1}}
\end{figure}
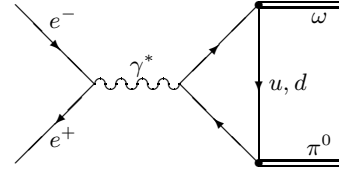

\begin{figure}
\begin{center}
\special{em:linewidth 0.6pt}
\linethickness{0.6pt}
\begin{picture}(150.00,80.00)
\put(40.00,40.00){\vector(-1,-1){15.0}}
\put(25.00,25.00){\line(-1,-1){15.0}}
\put(28.00,20.00){\makebox(0,0)[cc]{$e^+$}}
\put(10.00,70.00){\vector(1,-1){17.0}}
\put(25.00,55.00){\line(1,-1){15.0}}
\put(28.00,65.00){\makebox(0,0)[cc]{$e^-$}}
\put(42.00,40.00){\oval(4.00,4.00)[t]}
\put(46.00,40.00){\oval(4.00,4.00)[b]}
\put(50.00,40.00){\oval(4.00,4.00)[t]}
\put(54.00,40.00){\oval(4.00,4.00)[b]}
\put(48.00,48.00){\makebox(0,0)[cc]{$\gamma^*$}}
\put(64.00,40.00){\circle{16.00}}
\put(65.00,48.00){\vector(+1,0){1.0}}
\put(63.00,32.00){\vector(-1,0){1.0}}
\put(72.00,40.00){\circle*{3.00}}
\put(72.00,41.00){\line(+1,+0){20.0}}
\put(72.00,39.00){\line(+1,+0){20.0}}
\put(85.00,48.00){\makebox(0,0)[cc]{$\rho^0$}}
\put(92.00,40.00){\circle*{3.00}}
\put(92.00,40.00){\vector(+1,+1){15.0}}
\put(107.00,55.00){\line(+1,+1){15.0}}
\put(122.00,70.00){\vector(0,-1){33.0}}
\put(133.00,40.00){\makebox(0,0)[cc]{$u,d$}}
\put(122.00,40.00){\line(0,-1){30.0}}
\put(122.00,10.00){\vector(-1,+1){17.0}}
\put(107.00,25.00){\line(-1,+1){15.0}}
\put(122.00,11.00){\line(1,0){30.0}}
\put(122.00,09.00){\line(1,0){30.0}}
\put(145.00,65.00){\makebox(0,0)[cc]{$\omega$}}
\put(122.00,71.00){\line(1,0){30.0}}
\put(122.00,69.00){\line(1,0){30.0}}
\put(145.00,17.00){\makebox(0,0)[cc]{$\pi^0$}}
\put(122.00,70.00){\circle*{3.00}}
\put(122.00,10.00){\circle*{3.00}}
\end{picture}
\end{center}
\caption{Feynman diagram with $\rho$ meson exchange.
\label{Fig:2}}
\end{figure}

\begin{figure}
\begin{center}
\special{em:linewidth 0.6pt}
\linethickness{0.6pt}
\begin{picture}(150.00,80.00)
\put(40.00,40.00){\vector(-1,-1){15.0}}
\put(25.00,25.00){\line(-1,-1){15.0}}
\put(28.00,20.00){\makebox(0,0)[cc]{$e^+$}}
\put(10.00,70.00){\vector(1,-1){17.0}}
\put(25.00,55.00){\line(1,-1){15.0}}
\put(28.00,65.00){\makebox(0,0)[cc]{$e^-$}}
\put(42.00,40.00){\oval(4.00,4.00)[t]}
\put(46.00,40.00){\oval(4.00,4.00)[b]}
\put(50.00,40.00){\oval(4.00,4.00)[t]}
\put(54.00,40.00){\oval(4.00,4.00)[b]}
\put(48.00,48.00){\makebox(0,0)[cc]{$\gamma^*$}}
\put(64.00,40.00){\circle{16.00}}
\put(65.00,48.00){\vector(+1,0){1.0}}
\put(63.00,32.00){\vector(-1,0){1.0}}
\put(72.00,40.00){\circle*{6.00}}
\put(72.00,41.00){\line(+1,+0){20.0}}
\put(72.00,39.00){\line(+1,+0){20.0}}
\put(85.00,48.00){\makebox(0,0)[cc]{$\rho'$}}
\put(92.00,40.00){\circle*{6.00}}
\put(92.00,40.00){\vector(+1,+1){15.0}}
\put(107.00,55.00){\line(+1,+1){15.0}}
\put(122.00,70.00){\vector(0,-1){33.0}}
\put(133.00,40.00){\makebox(0,0)[cc]{$u,d$}}
\put(122.00,40.00){\line(0,-1){30.0}}
\put(122.00,10.00){\vector(-1,+1){17.0}}
\put(107.00,25.00){\line(-1,+1){15.0}}
\put(122.00,11.00){\line(1,0){30.0}}
\put(122.00,09.00){\line(1,0){30.0}}
\put(145.00,65.00){\makebox(0,0)[cc]{$\omega$}}
\put(122.00,71.00){\line(1,0){30.0}}
\put(122.00,69.00){\line(1,0){30.0}}
\put(145.00,17.00){\makebox(0,0)[cc]{$\pi^0$}}
\put(122.00,70.00){\circle*{6.00}}
\put(122.00,10.00){\circle*{6.00}}
\end{picture}
\end{center}
\caption{Feynman diagram with $\rho'$ meson exchange.
\label{Fig:3}}
\end{figure}

For the description of the first two diagrams, see Figs.~\ref{Fig:1} and \ref{Fig:2},
we need the part of the standard NJL Lagrangian which describes interactions
of photons, pions and vector $\rho$ and $\omega$
mesons with quarks, see Refs.~\cite{Volkov:1986zb,Volkov:1993jw,Ebert:1994mf}. It has the form
\bea \label{L1}
\Delta{\mathcal L}_{1} &=& \bar{q}\biggl[i\hat\partial - m + eQ\hat{A}
+ig_{\pi}\gamma_5\tau_3\pi^0
\nonumber \\
&+& \frac{g_\rho}{2}\gamma_\mu\left(I\hat{\omega} + \tau_3\hat{\rho}^0\right) \biggr]q,
\eea
where $\bar{q}=(\bar{u},\bar{d})$ with $u$ and $d$ quark fields;
$m=diag(m_u,m_d)$, $m_u=m_d=280$~MeV is the constituent quark mass;
$Q=diag(2/3,-1/3)$ is the electromagnetic quark charge matrix;
$e$ is the electron charge;
$A$, $\pi^0$, $\omega$ and $\rho^0$ are the photon, pion, $\omega$ and $\rho$ meson fields, respectively;
$g_\pi$ is the pion coupling constant,
$g_\pi=m_u/f_\pi$, where $f_\pi=93$~MeV is the pion decay constant;
$g_\rho$ is the vector meson coupling constant, $g_\rho\approx 6.14$ corresponding
to the standard relation $g_\rho^2/(4\pi)=3$;
$I=diag(1,1)$ and $\tau_3$ is the third Pauli matrix.

All three amplitudes contain the common part corresponding to $e^+e^-\gamma$ vertex and
the photon propagator. So the sum of the amplitudes can be cast in the form
\bea\label{Tlambda}
T^{\lambda} = \bar{e}\gamma_\mu e \frac{1}{s}
\bigl\{T_1^{\mu\lambda}+T_2^{\mu\lambda}+T_3^{\mu\lambda}\bigr\}\varepsilon_{\lambda}(\omega),
\eea
where $s=(p_1(e^+) + p_2(e^-))^2 \equiv q^2$.
The first part $T_1$ is just the triangle quark diagram of the anomaly type.
Note that the loop integral in it is finite. Following
papers~\cite{Bystritskiy:2007wq,Bystritskiy:2009zz,Arbuzov:2010vq},
we use here the na\"ive confinement approach and neglect the imaginary part of
the loop integral.

The integral over the energy $k^0$ of the virtual loop momentum is calculated
analytically using the residue method. The integral over $\vec{k}$ is taken
numerically. Even so that this integral is convergent, we put the cut off
for the upper  value of $|\vec{k}|$ equal $\Lambda=1.03$~GeV~\cite{Ebert:1992ag}.
This cut off will be necessary in contributions of the radially excited mesons states.
Here the cut off is applied for homogeneity of the approach. And
the numerical result for the convergent integral do not change considerably if
the cut off would be removed.
The imaginary part is neglected by taking the principal value of the integral.

The second contribution $T_2$ contains three factors. The first one is the
transition of photon into $\rho$ meson which is described in Ref.~\cite{Volkov:1986zb}:
\bea \label{gamma_rho}
\frac{e}{g_\rho} \bigl( g^{\nu\nu'}q^2 - q^{\nu}q^{\nu'}\bigr).
\eea
Note that contrary to the case of the triangle diagram, the quark loop describing the
$\gamma-\rho$ transition contains a logarithmic divergence. The standard NJL methods
were applied for its regularization using the cut off value.
The second factor is the $\rho$ meson propagator,
\bea
\frac{ig^{\nu'\nu''}}{q^2-M_\rho^2+iM_\rho\Gamma_\rho}\, ,
\eea
where the neutral $\rho$-meson mass $M_\rho=775$~MeV and width
$\Gamma_\rho=146$~MeV~\cite{Nakamura:2010zzi}.
Note that the non-diagonal terms in the numerator of the vector particle
propagator were dropped because of the gradient invariance of the triangle diagram.
The third factor is the same triangle diagram as in the first amplitude $T_1$.

A more complicated situation appear for the third contributions $T_3$, see Fig.~\ref{Fig:3}, because
we deal here with radially excited $\rho'$ meson. Instead of the Lagrangian~(\ref{L1})
we use here an extended version of the NJL Lagrangian which allows us to describe
both ground and radial-excited meson states~\cite{Volkov:1996fk,Volkov:1997dd,Arbuzov:2010vq}:
\bea \label{L2}
\Delta {\mathcal L}_2 &=&
\bar{q}(k')\biggl\{i\hat\partial - m + eQ\hat{A}
+ A_\pi \tau^3\gamma_5\pi^0(p)
+ A_\omega \hat{\omega}(p)
\nn \\
&-& A_{\rho'} \tau^3{\hat{\rho^0}'}(p)
\biggr\}q(k), \ \ \ p = k-k',
 \\ \nn
A_\pi &=& g_{\pi_1}\frac{\sin(\alpha+\alpha_0)}{\sin(2\alpha_0)}
       +g_{\pi_2}f({k^\bot}^2)\frac{\sin(\alpha-\alpha_0)}{\sin(2\alpha_0)},
\nn \\
A_\omega &=& g_{\rho_1}\frac{\sin(\beta+\beta_0)}{\sin(2\beta_0)}
       +g_{\rho_2}f({k^\bot}^2)\frac{\sin(\beta-\beta_0)}{\sin(2\beta_0)},
\nn \\ \nn
A_{\rho'} &=& g_{\rho_1}\frac{\cos(\beta+\beta_0)}{\sin(2\beta_0)}
        +g_{\rho_2}f({k^\bot}^2)\frac{\cos(\beta-\beta_0)}{\sin(2\beta_0)}.
\eea
The radially-excited states were introduced in the NJL model with the help of
the form factor in the quark-meson interaction:
\bea
f({k^\bot}^2) &=& (1-d |{k^\bot}^2|) \Theta(\Lambda^2-|{k^\bot}^2|),
\nn \\
{k^\bot} &=& k - \frac{(kp)p}{p^2},\qquad d = 1.78\ {\mathrm{GeV}}^{-2},
\eea
where $k$ and $p$ are the quark and meson momenta, respectively.
The filled circles in Fig.~\ref{Fig:3} denote the presence of the form
factor in the quark--meson vertexes.
Note that the NJL model itself and its extended version
can be used only for sufficiently low energies. In this
study we attempt to receive qualitative results working at energies up to 2~GeV.

Coupling constants $g_{\pi_1}$ and $g_{\rho_1}$ coincide with $g_\pi$ and $g_\rho$ constants
introduced above in the standard NJL version.
The other coupling constants are defined via one-loop integrals:
\bea
g_{\pi_2} = \left[4I_2^{f^2}\right]^{-1/2},
\qquad
g_{\rho_2} = \left[\frac{2}{3}I_2^{f^2}\right]^{-1/2}=\sqrt{6}g_{\pi_2},
\eea
where
\bea \nn
I_m^{f^n} &=& - i N_c \int \frac{\dd^4k}{(2\pi)^4}
\frac{\bigl(f({k^\bot}^2)\bigr)^n}{(m^2 - k^2)^m}, \quad n,m=1,2.
\eea
The angles $\alpha_0=59.06^\circ$, $\alpha=59.38^\circ$, $\beta_0=61.53^\circ$ and $\beta=76.78^\circ$
were defined in Ref.~\cite{Volkov:1997dd,Arbuzov:2010vq} to describe mixing of the ground and
excited meson states. This contribution $T_3$ again consists of 3 parts.
The $\gamma-\rho_2$ transition ($\gamma-\rho_1$ transition coincides with the standard $\gamma-\rho$ one)
can be expressed via the $\gamma-\rho$ transition~(\ref{gamma_rho}) with the additional
factor~\cite{Volkov:1996fk,Volkov:1997dd}
\bea
\Gamma = \frac{I_2^f}{\sqrt{I_2I_2^{f^2}}}\approx 0.47.
\eea
So the $\gamma-\rho'$ transition takes the form
\bea \nn
\frac{e}{g_\rho} \bigl( g^{\nu\nu'}q^2 - q^{\nu}q^{\nu'}\bigr)\biggl\{
\frac{\sin(\beta+\beta_0)}{\sin(2\beta_0)} + \Gamma \frac{\sin(\beta-\beta_0)}{\sin(2\beta_0)}
\biggr\}.
\eea
We take the $\rho'$ propagator is taken in the Breit-Wigner form
\bea
\frac{g^{\nu'\nu''}}{q^2-M^2_{\rho'}+i\sqrt{q^2}\Gamma_{\rho'}(q^2)}\, ,
\eea
where the running $\rho'$ width reads
\bea
&& \Gamma_{\rho'}(q^2) = \Gamma(\rho'\to2\pi) + \Gamma(\rho'\to\omega\pi^0)
+  (\Gamma_{\rho'}(M_{\rho'})
\nonumber \\ && \quad 
-\Gamma(\rho'\to\omega\pi^0)-\Gamma(\rho'\to\omega\pi^0))
\Theta(\sqrt{s}-M_{a_1}+M_\pi)
\nonumber \\ && \quad \times
\biggl(\frac{p_{a_1}(s)}{p_{a_1}(M_{\rho'})}\biggr),
\eea
where $p_{a_1}(s)$ is the momentum of $a_1$ meson in the decay $\rho'\to a_1\pi$.
We assume that below the threshold of the reaction $\rho'\to a_1\pi$ the main
contribution the the width is given by the two channels  $\rho'\to2\pi$
$\rho'\to\omega\pi^0$. Above the peak $\sqrt{s}\geq M_{\rho'}$,
where many other channels are opened, we use the complete width $\Gamma_{\rho'}=340$~MeV
(we take the value at the lower PDG~\cite{Nakamura:2010zzi} boundary).
The transition to the complete width is approximately described by linear
switching on of the contribution due to the decay $\rho'\to a_1\pi$ 
being one of the most probable channels.
The values $\Gamma(\rho'\to2\pi)=22$~MeV
and $\Gamma(\rho'\to\omega\pi^0)=75$~MeV were calculated in \cite{Volkov:1997dd}
in agreement with the experimental data~\cite{Clegg:1993mt}.
Since we are working close to the $\omega\pi$ threshold, taking it into account in the
running width is important. Running of the $\rho$ meson width is less important numerically, since
the $\rho$ meson contribution is relatively small.

\section{Numerical Results and Discussion}

Now we can estimate the contributions of the considered amplitudes into the total
process cross section. The details of phase volume calculations and evaluation of
the cross section can be found in Ref.~\cite{Bystritskiy:2009zz}. For our case it takes
the form
\bea \label{sigma}
&& \sigma(s) = \frac{3\alpha^2}{32\pi^3s^3}\lambda^{3/2}(s,M_\omega^2,M_\pi^2)
\frac{g_\rho^2}{f_\pi^2}|J^{(3)}|^2 
\nn \\ && \quad \times
{\mathrm Br}(\omega\to\pi^0\gamma),
\\ \nn
&& \lambda(s,M_\omega^2,M_\pi^2) = (s-M_\omega^2-M_\pi^2)^2-4M_{\omega}^2M_{\pi}^2,
\eea
where
\bea \label{J3}
&& J^{(3)} = \biggl(1 - \frac{q^2}{q^2-M_\rho^2+iM_\rho\Gamma_\rho}\biggr)I^{(3)}_\gamma
\nn \\
&& \quad + \frac{\Gamma q^2}{q^2-M_{\rho'}^2+i\sqrt{q^2}\Gamma_{\rho'}(q^2)}I^{(3)}_{\rho'},
\eea
and
\bea \nn
&& I^{(3)}_\gamma\left(\frac{m^2}{s}\right) = \int\frac{\dd^4k}{i\pi^2}
\frac{m^2 \Theta(\Lambda^2-|{k^\bot}^2|)}
{(k^2-m^2+i0)}
\\ \nn
&&\quad \times \frac{1}
{((k-P_\omega)^2-m^2+i0)((q-k)^2-m^2+i0)}.
\eea
In the first line of Eq.~(\ref{J3}) we have the sum of the photon and rho meson exchange
contributions. Their sum takes the form that coincides with the one received in the
vector meson dominance model, see {\it e.g.}~\cite{Akhmetshin:2003ag}. 
In fact, the standard NJL model contains the vector dominance
model~\cite{Volkov:1986zb,Volkov:1982zx,Ebert:1982pk}.

Note that keeping the cut-off for the convergent integral in $I^{(3)}_\gamma$ entering $T_1$
and $T_2$ in Eq.~(\ref{Tlambda}) is not necessary, but it does not affect much the numerical result.
Expression for the integral $I^{(3)}_{\rho'}$ has a rather cumbersome form and contains a combination
of terms with different powers of the form factor (up to the third power). It is constructed according
to the Feynman rules coming from the Lagrangian~(\ref{L2}). 
For calculation of the relevant quark loop integrals we use the method described in
Ref.~\cite{Arbuzov:2010vq}. It is worth to note that in 
our calculations the signs of $I^{(3)}_{\gamma}$ and $I^{(3)}_{\rho'}$ appeared to  be opposite 
in accordance with the fit to experimental data performed in~\cite{Akhmetshin:2003ag}. 

The coupling constants $g_\rho=6$ and $f_\pi=93$~MeV entering in Eq.~\ref{sigma} are universal input 
parameters for the NJL model. 
In Ref.~\cite{Akhmetshin:2003ag} another value for this constant was used: $f_\rho\approx 5$ received
from the decay width $\Gamma(\rho\to e^+e^-)$. Another difference is coming from the value
for the coupling constant in the vertex $\rho\omega\pi$. In our model it is  
$g_{\rho\omega\pi}=3g_\rho^2/(8\pi^2f_\pi)\approx 14.7$~GeV$^{-1}$, while in Ref.~\cite{Akhmetshin:2003ag}
the value $g_{\rho\omega\pi}\approx 17$~GeV$^{-1}$ taken as a fitting parameter.

\begin{center}
\begin{figure}
\includegraphics[width=6.2cm,angle=270]{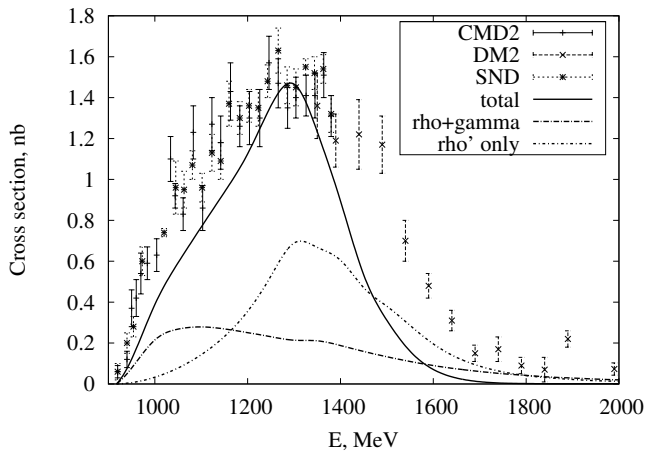}
\caption{Comparison of experimental results for $e^+e^-\to\pi^0\omega\to\pi^0\pi^0\gamma$
with the NJL model prediction (lines).
\label{plot}}
\end{figure}
\end{center}

Fig.~\ref{plot} shows the experimental data~\cite{Bisello:1990du,Achasov:2000wy,Akhmetshin:2003ag}
and the corresponding theoretical prediction (the solid line) receive within the applied here NJL
phenomenological model.
The dash-dotted line shows the sum of the photon and rho-meson exchange contributions.
The short-dash-dotted corresponds to the pure $\rho'$ meson exchange.
The photon and $\rho$ meson exchange is important for the threshold region, while
the $\rho'$ contribution dominates in the region $\sqrt{s}\sim M_{\rho'}$.
Note that the NJL model is adjusted for applications at low energies up to about 2~GeV.
In this energy range,
the model gives a qualitative description of meson properties and interactions. The
advantage is that the set of parameters is limited and fixed.
Note that to describe the given process we did not introduce any new parameter in the model.
Presumably, adding of the $\rho''(1700)$ meson contribution might improve the agreement with 
the experimental data above the peak, but for the time being the NJL model is not suited 
to include the second radial excitations of mesons with large masses. 
 A more accurate description of the
threshold behavior requires going beyond the Hartree-Fock approximation that was
used here.  Indeed, meson-meson final state interactions can play an important role
in the threshold domain. 
%Moreover as one can test from numerical evaluations,
%relative phases between different channels are important.

The same approach was successfully applied in papers~\cite{Volkov:1997dd,Volkov:2001ct,Volkov:1999yi}
for description of mass spectra and strong decays with participation of excited mesons.
In the present work we continue the work started in Refs.~\cite{Kuraev:2009uh,Arbuzov:2010vq} devoted
to description of radiative decays with participation of radially excited mesons and
pass to description of annihilation processes studied at modern $e^+e^-$ colliders.
Similar mechanism appear in the processes of $e^+e^-$ annihilation into {\it e.g.} $\pi^0\gamma$,
$\pi'\gamma$, and ${\pi'}^\pm\pi^\mp$ which will be considered elsewhere.

\subsection*{Acknowledgments}
We are grateful to A.~Akhmedov, S.~Gerasimov, and
G.~Fedotovich for useful discussions.
This work was supported by RFBR grant 10-02-01295-a.

\end{document}